\documentclass[a4paper,12pt]{article}%
\usepackage{jheppub}
\usepackage{amsmath,amsfonts,setspace,cancel,url,hyperref,verbatim,slashed,amsthm,graphicx,color,soul,lipsum}
\pdfoutput=1
\usepackage[utf8]{inputenc}
\newcommand{\be}{\begin{equation}}
\newcommand{\ee}{\end{equation}}
\usepackage{accents}
\newcommand\starred[1]{\accentset{\star}{#1}}
\newcommand\im{{\rm im}\: }

\newcommand\lf{$L_\infty$}

\begin{document}


\title{The $L_\infty$-algebra of the S-matrix}

\author{Alex S. Arvanitakis}
\affiliation{\small \em The Blackett Laboratory,\\
\small \em Imperial College London, \\
\small \em Prince Consort Road London SW7 2AZ, U.K.}
\emailAdd{a.arvanitakis@imperial.ac.uk}
\abstract{
We point out that the one-particle-irreducible vacuum correlation functions of a QFT are the structure constants of an $L_\infty$-algebra, whose Jacobi identities hold whenever there are no local gauge anomalies.  The LSZ prescription for S-matrix elements is identified as an instance of the ``minimal model theorem'' of $L_\infty$-algebras. This generalises the algebraic structure of closed string field theory to arbitrary QFTs with a mass gap and leads to recursion relations for amplitudes (albeit ones only immediately useful at tree-level, where they reduce to Berends-Giele-style relations as shown in \cite{Macrelli:2019afx}).

}

\preprint{Imperial-TP-2019-ASA-01}



\titlepage

\maketitle


\section{Introduction}

Recall the setting of ``textbook'' quantum field theory (QFT) on Minkowski spacetime. The first goal therein is to calculate scattering amplitudes. These are obtained from correlation functions through the \emph{LSZ reduction formula} \cite{Lehmann:1954rq}
\be
\label{lsz}
\begin{split}
(i/\sqrt{Z})^n\int \prod_{i=1}^n dx_i\:\big[ \exp(i x_i\cdot p_i)(\square_{x_i}+m^2)\langle\phi(x_1)\phi(x_2)\cdots \phi(x_n)\rangle_\text{conn.}\big]
\end{split}
\ee
stated here for a connected S-matrix element (see \cite{Itzykson:1980rh} section 5-1-5) involving $n\geq 3$ outgoing spin-0 particles of the same species, mass $m^2>0$ and on-shell momenta $p_1,p_2,\dots p_n$ respectively. ($\phi(x)$ is the field operator with nonzero amplitude $Z$ to create a 1-particle state while acting on the vacuum $|0\rangle$, $\langle\phi(x_1)\phi(x_2)\cdots \phi(x_n)\rangle_\text{conn.}$ is its connected $n$-point time-ordered vacuum correlation function, and $(\square_{x_i}+m^2)$ is the Klein-Gordon operator, acting on the correlator only.) One is then concerned with the calculation of correlators from first principles.

In this paper we make the following observations.

Firstly, that implicit in Zinn--Justin's 1974 proof of perturbative renormalisability of gauge theories \cite{ZinnJustin:1974mc} is an $L_\infty$-algebra (a generalisation of a Lie algebra involving brackets of any arity), whose structure constants are the one-particle-irreducible (1PI) correlators. We shall see that the $L_\infty$ Jacobi identities are equivalent to the Zinn--Justin equation, i.e.~the classical Batalin-Vilkovisky (BV) \cite{Batalin:1984jr,Batalin:1985qj,Batalin:1984ss,Batalin:1981jr,Batalin:1977pb} master equation for the 1PI generating functional $\Gamma[\Phi,\starred{\Phi}]$ (including sources $\starred \Phi$ for BRST transformations), which is in turn equivalent to the absence of anomalous terms in Slavnov-Taylor identities. We call this the \emph{$L_\infty$-algebra of correlators}.

Secondly, and less trivially, that the LSZ reduction formula endows the asymptotic 1-particle states with the structure of a \emph{minimal} $L_\infty$-algebra (one without unary bracket), whose structure constants are the connected S-matrix elements. The $L_\infty$-algebraic interpretation is that this \emph{$L_\infty$-algebra of the S-matrix} is a \emph{minimal model} (no relation to the CFT notion) for the aforementioned $L_\infty$-algebra of 1PI correlators, and that the LSZ formula defines the corresponding \emph{quasi-isomorphism}. We will explicitly show this for scalar theories with a mass gap only, but the argument directly generalises to any gapped theory.

The definition of $L_\infty$-algebras \cite{Zwiebach:1992ie,Lada:1992wc} is now overdue. They generalise Lie algebras, so besides the structure constants $C^a_{bc}$ defining a binary bracket, one has $C^a_b,\, C^a_{b_1b_2b_3},\dots C^a_{b_1\dots b_n},\dots$ defining a unary, ternary, \dots $n$-ary\dots ~ bracket respectively, obeying symmetry identities and Jacobi identities (see \eqref{Qdef}). Their first explicit appearance\footnote{They have appeared implicitly earlier in the D'Auria-Fr\'e formulation of supergravity \cite{DAuria:1982uck,DAuria:1980cmy,Castellani:1982kd} and recognised as such in \cite{Sati:2008eg}; later they also appeared again implicitly in the work of \cite{Berends:1984rq} on higher-spin particles. (Both of these appearances predate the CSFT observation and I am grateful to Jim Stasheff for pointing this out.)} was in Zwiebach's work on closed bosonic string field theory \cite{Zwiebach:1992ie} (CSFT): the genus-zero closed string correlators are the structure constants of an $L_\infty$-algebra, whose Jacobi identities imply gauge invariance of the string field lagrangian. Since then the concept has been picked up by mathematicians, who have articulated a general philosophy \cite{schlessinger2012deformation}
: deformations of a structure (e.g. the complex structure of a complex manifold) are solutions of the Maurer-Cartan equation \eqref{MaurerCartan} (resp. the Kodaira-Spencer equation) of the associated $L_\infty$-algebra. For CSFT this is the string field equation of motion, whose solutions were argued to determine the conformal manifold of the worldsheet CFT \cite{sen1990equations}. For our $L_\infty$-algebra of correlators, this lore boils down to the best-known application of the 1PI functional $\Gamma$ in its guise as the Coleman-Weinberg effective potential \cite{Coleman:1973jx}: extrema of $\Gamma$ determine vacuum states.

Minimal $L_\infty$-algebras have $C^a_{b}=0$. The relation between minimal $L_\infty$-algebras and S-matrices has been anticipated, originally in the context of 2D string theory by Witten and Zwiebach \cite{Witten:1992yj} and Verlinde \cite{Verlinde:1992qa}, later for more general string (field) theories by authors including Kajiura \cite{kajiura}, M\"unster and Sachs \cite{Munster:2012gy}, and Konopka \cite{Konopka:2015tta}, and most recently for tree-level gravity and Yang-Mills by N\"utzi and Reiterer \cite{Nutzi:2018vkl}. The novelty in our work is the generalisation beyond both string theory and perturbative expansions. To this end we write a proof of the \emph{minimal model theorem} (that takes an $L_\infty$-algebra and gives the canonically associated minimal one), following a suggestion in \cite{kajiura}, that mirrors the derivation by Jevicki and Lee \cite{Jevicki:1987ax} of the S-matrix generating functional from the 1PI functional $\Gamma$.

The recent resurgence of physics interest in $L_\infty$-algebras (e.g. \cite{Hohm:2017pnh,Jurco:2018sby,Ritter:2015ymv,Fiorenza:2011jr,Kotov:2007nr,Hohm:2017cey,Blumenhagen:2017ogh,Cederwall:2018aab,Cagnacci:2018buk,Lavau:2014iva,Lavau:2017tvi,Arvanitakis:2018cyo}) mostly centres on gauge symmetries of classical theories. (Most relevant here is \cite{Jurco:2018sby}, which articulates the lore that classical BV master actions have canonical associated $L_\infty$-algebras \cite{alexandrov1997geometry,Fisch:1989rp,Barnich:1997ij,Fulp:2002kk,Berends:1984rq,Movshev:2003ib,Movshev:2004aw,Zeitlin:2007vv,Zeitlin:2007vd,Zeitlin:2007yf,Zeitlin:2007fp,Zeitlin:2008cc,Rocek:2017xsj}). Our observations suggest that $L_\infty$-algebraic approaches might be even more natural for quantum field theories: in a sense, $L_\infty$-algebras have been underlying QFT all along, as corroborated by (what we will argue is) the natural algebraic connection between vacuum correlators and Minkowski space S-matrix elements. We will suggest generalisations to other backgrounds and QFTs in the Discussion.

\bigskip

{\bf Note added:} The first version of this paper originally appeared on the arXiv at the 
same time as \cite{Macrelli:2019afx} which makes the same points (for tree-level theories) 
and in addition proves the relation between recursive amplitude formulae 
and the minimal model theorem.
In particular, that is where the quasi-isomorphism appearing in the proof of the minimal model theorem for the \lf{}-algebra of correlators was originally given a physical interpretation as the collection of Berends-Giele ``off-shell currents'' \cite{Berends:1987me} appearing in their namesake recursion relations.


\section{$L_\infty$-algebras}
 An $L_\infty$-algebra structure lives on a $\mathbb Z$-graded vector space $\mathcal V$. Denoting the $L_\infty$-algebra generators by $T_a$, each assumed to have some definite ``$L_\infty$-degree'' $\deg T_a\in\mathbb Z$, the symmetry and Jacobi identities are together encoded in the BRST-charge-like operator $Q$ (where $\partial/\partial z^a=\partial_a$ is a left derivative; our notation for symplectic supermanifolds is as in \cite{Arvanitakis:2018cyo} Appendix~A)
\be
\label{Qdef}
Q=\sum_{n=1}^\infty \frac{1}{n!}C^a_{b_1\dots b_n} z^{b_1}\cdots z^{b_n}\frac{\partial}{\partial z^a}=Q^a\partial_a\,,\qquad Q^2=0\,, \quad \deg Q=+1\,.
\ee
$Q$ acts on the space $C(\mathcal V)$ of formal power series in variables $z^a$, with multiplication
\be
z^a z^b=(-1)^{(\deg z^a)(\deg z^b)}z^b z^a=(-1)^{ab}z^b z^a\,,
\ee
so $z^a$ is bosonic or fermionic according to $\deg z^a \in\mathbb Z$. The structure constants $C^a_{b_1\dots b_n}$ (real or complex numbers) are defined so $Q$ increases this degree by 1. The collection $\{z^a\}$ is the dual basis to the generators $T_a$ and the $z^a$ inherit their degrees from the $T_a$: $\deg z^a=-\deg T_a$. A Lie algebra is the special case where all $\deg z^a=1$; $Q^2=0$ then reduces to the familiar Jacobi identity. The $L_\infty$ brackets are defined by
\be
[T_{b_1},T_{b_2},\dots T_{b_n}]=C^a_{b_1\dots b_n} T_a\,,\quad \deg [T_{b_1},T_{b_2},\dots T_{b_n}]=1+\sum \deg T_{b_i}
\ee
(This $L_\infty$ degree convention agrees with the ``target space ghost number'' of \cite{Zwiebach:1992ie} and the ``$b$-picture'' of \cite{Hohm:2017pnh} after $\deg T_a\to -\deg T_a$.)

We exclusively use this geometric definition (called the ``DGA-picture'' in \cite{Jurco:2018sby}) where $Q$ is interpreted as a vector field on a formal superspace \cite{Alexandrov:1995kv,kontsevich2003deformation} with coordinates $z^a$, whose ring of ``functions'' is $C(\mathcal V)$ by definition. A \emph{homomorphism} $f$ of $L_\infty$-algebras $f: \mathcal V\to \mathcal V'$ (henceforth \emph{morphism}) is a degree 0 map of superspaces of the form (where each $f^{\phantom{n}a'}_{n\,a_1a_2\dots a_n}$ is a constant)
\be
\label{morphism}
f^\star (z')^{a'}\equiv f^{\phantom{1}a'}_{1\,a} z^a+ \frac{1}{2} f^{\phantom{2}a'}_{2\,a_1a_2}z^{a_1}z^{a_2}+\dots\equiv \sum_{n=1}^{\infty} \frac{1}{n!}f^{\phantom{n}a'}_{n\,a_1a_2\dots a_n}z^{a_1}z^{a_2}\cdots z^{a_n}
\ee
which relates the vector fields by
\be
\label{morphismcondition}
Q\circ f^\star=f^\star \circ Q'\,.
\ee
More precisely it is an $L_\infty$-degree-preserving linear map of the spaces of polynomials $f^\star:C(\mathcal V')\to C(\mathcal V)$ which respects multiplication ($f^\star(ab)=f^\star(a)f^\star(b)$) and has zero constant part. It is an isomorphism if $f_1$ is invertible as a map $\mathcal V\to \mathcal V'$.

An $L_\infty$-algebra with invariant inner product $\kappa$ is called \emph{cyclic}\footnote{The name is in reference to the $A_\infty$ generalisation: an $A_\infty$-algebra is defined exactly as above with the exception that the product $z^a z^b$ is only associative instead of graded commutative. Then a degree $-1$ symplectic form $\kappa$ is annihilated by the corresponding $Q$ iff the index-down structure constants have a \emph{cyclic} symmetry under permutations. In terms of the relation between open string field theory and $A_\infty$-algebras, this corresponds to the fact that open string vertex operators are inserted on the $S^1$ boundary of the worldsheet.}. This is defined as a degree $-1$ symplectic form $\kappa=\kappa_{ab}dz^a dz^b/2$ with constant coefficients $\kappa_{ab}$, annihilated by the Lie derivative $\mathcal L_Q$. The last condition is equivalent to
\be
\label{cyclicity}
C_{b_1\dots b_{n+1}}\equiv\kappa_{ab_1}C^a_{b_2\dots b_{n+1}}=(-1)^{b_1b_2} \kappa_{ab_2}C^a_{b_1\dots b_{n+1}}\,.
\ee
This notion agrees with the inner product on the space of string states of Zwiebach's \cite{Zwiebach:1992ie} (cf.~the ``multilinear string functions'' therein). $\kappa_{ab}$ has the symmetry
\be
\kappa_{ab}=\kappa_{ba}\,,
\ee
so the inverse $\kappa^{ab}$ is also symmetric. (The sign factor $(-1)^{(a+1)(b+1)}$ evaluates to $+1$.)

Cyclic $L_\infty$-algebras are related to the BV formalism: the formal power series $\Theta$
\be
\label{thetadef}
\Theta(z)\equiv\sum_{n=2}^\infty \frac{1}{n!}C_{b_1\dots b_n} z^{b_1}\cdots z^{b_n}\,,\quad \deg \Theta=0
\ee
defines a cyclic $L_\infty$-algebra with
\be
Q\equiv(\Theta,\--)=\partial_b\Theta \kappa^{ba}\partial_a
\ee
when $(\Theta,\Theta)=0$ (where $Q$ is the hamiltonian vector field of $\Theta$, and $(\--,\--)$ is the Poisson bracket of $\kappa$), and vice versa. By introducing ``antifields'' $\starred z_a$ we can replace an arbitrary $L_\infty$-algebra $\mathcal V$ with a cyclic one $\mathcal V_\text{cyclic}$ (whose $\Theta$ is \eqref{Qdef} with $\partial_a\to \starred z_a$), yielding a surjective morphism $\mathcal V_\text{cyclic}\to\mathcal V$. (This is the ``odd double'' of \cite{braun2015unimodular}.)

For completeness we mention the Maurer-Cartan equation for a cyclic $L_\infty$-algebra. Let $\Psi=\Psi^a T_a$ be a degree-zero element of the vector space $\mathcal V$ of a cyclic $L_\infty$-algebra (with coordinates $\Psi^a$). Consider the translation $z^a\to z^a+\Psi^a$ generated by the vector field $v_\Psi\equiv\Psi^a\partial_a$. Since the symplectic form $\kappa$ has constant coefficients,
\be
0=\exp(\mathcal L_{v_\Psi}) \big(\Theta(z),\Theta(z)\big)=\big(\Theta(z+\Psi),\Theta(z+\Psi)\big)\,.
\ee
where $\Theta(z+\Psi)\equiv \exp(\mathcal L_{v_\Psi}) \Theta(z)$. Therefore, $\Theta(z+\Psi)$ is a formal power series in $z^a$ which will define a cyclic $L_\infty$-algebra iff the linear in $z$ term below vanishes:
\be
\Theta(z+\Psi)=\Theta(\Psi)+ z^a \left(\frac{\partial}{\partial z^a}\Theta\right)_{z=\Psi} + O(z^2)\,.
\ee
By degree-counting, this is true iff $\Psi$ solves the \emph{Maurer-Cartan equation}
\be
\label{MaurerCartan}
\frac{\partial\Theta(\Psi)}{\partial \Psi^a}=0\,.
\ee
Note that there are convergence issues here because the $\Psi^a$ are reals (if the $L_\infty$-algebra is real) so $\Theta(\Psi)$ is an honest power series as opposed to a formal one.


Now consider the Jacobi identities $Q^2=0$ of an arbitrary $L_\infty$-algebra. These are organised by \emph{polynomial} degree in $z^a$: split $Q=Q_0+Q_1+Q_2+\dots$ so each $Q_n$ increases polynomial degree by $n$, to obtain infinitely many identities
\be
Q_0^2=0\,,\quad Q_0Q_1+Q_1Q_0=0\,,\quad  Q_1^2 + Q_2 Q_0 + Q_0 Q_2=0\,,\dots
\ee
The first gives $C^a_b C^b_c=0$; i.e.~the $L_\infty$-algebra unary bracket $K$ (defined so $K(T_a)=C_a^b T_b$) has $K^2=0$. Since $K$ raises $L_\infty$-degree by 1, $K$ is a cohomology operator on the graded vector space $\mathcal V$.

For a \emph{minimal} $L_\infty$-algebra (i.e.~$K=0$; in particular, Lie algebras are minimal as $L_\infty$ ones), the underlying vector space $\mathcal V$ is the cohomology of $K$. More generally, for any $L_\infty$-algebra one can put an $L_\infty$-algebra structure on the cohomology of $K$. This is called a \emph{minimal model} for the original algebra, and all minimal models thusly obtained are isomorphic. For most purposes the study of an $L_\infty$-algebra can be reduced to that of its minimal model; for this reason, morphisms of $L_\infty$-algebras which correspond to isomorphisms of minimal models are particularly important and are known as \emph{quasi-isomorphisms}. They are equivalently characterised as morphisms $f$ such that $f_1:\mathcal V\to V'$ is an isomorphism on the cohomologies of $K,K'$ respectively.

\subsection{The minimal model theorem for cyclic $L_\infty$-algebras}
The \emph{minimal model theorem} \cite{kadeishvili1980homology} claims a minimal $L_\infty$-algebra $\mathcal V_\text{min}$ and an injective quasi-isomorphism $\mathcal V_\text{min}\to\mathcal V$. We here provide a short construction of a minimal model for a cyclic $L_\infty$-algebra ($\mathcal V, \Theta(z), \kappa$) following a suggestion of Kajiura \cite{kajiura}: roughly, one extremises the hamiltonian $\Theta$, then backsubstitutes to find a hamiltonian for a minimal model. We will see later how this is exactly like the Jevicki-Lee prescription for the S-matrix \cite{Jevicki:1987ax}.


\begin{proof}
(valid when $({\mathcal V}^\star)^\star\cong \mathcal V$, e.g.~in finite dimensions.)

We invoke a ``cyclic Hodge-Kodaira decomposition''  (see \cite{Kajiura:2001ng,kajiura,Jurco:2018sby} and appendix A)
\be
\label{hodgekodaira}
\mathcal V=P\oplus P^\perp
\ee
where $P$ is a subspace of $K$-cohomology representatives, $P^\perp$ is its $\kappa$-orthogonal complement, and $\kappa$ restricted to either is non-degenerate. We can find a partial inverse $G$ of $K$ on $P^\perp$ (a degree $-1$ map we will call the \emph{propagator}), so $GKG=G,\,KGK=K,$ $G^2=0$ and $G_{ac}\equiv\kappa_{ab}G^{b}_c=-(-1)^{ac} G_{ca}$.  In particular $P^\perp=\im(KG)\oplus\im(GK)$ where each summand is $\kappa$-null. 
We therefore have
\be
\label{kappablockdiagonal}
\kappa=\begin{pmatrix}\kappa|_P &0\\ 0&\kappa|_{P^\perp} \end{pmatrix}\,,\quad \kappa|_{\im (KG)}=\kappa|_{\im (GK)}=0\,,\quad \quad\kappa|_P=P\kappa P \text{ non-degenerate.}
\ee

Now to construct a quasi-isomorphism $f:\mathcal V_\text{min}\equiv P\to \mathcal V$. Let $\zeta^a$ be a basis of $P^\star\subset \mathcal V^\star$. Extend it to a basis $z^a$ of $\mathcal V$ to write a direct sum
\be
z^a=\zeta^a+ G^a_b C^b_c z^c + C^a_b G^b_c z^c\,,\quad P^a_b \zeta^b=\zeta^a\,.
\ee
Here $P^a_b$ is the matrix of the projector $P$ onto cohomology representatives (also denoted $P$), and $C^a_b$ is the matrix of $K$. Specifying the last two terms as formal power series in $\zeta^a$ specifies the quasi-isomorphism (assuming both are $O(\zeta^2)$).

The candidate quasi-isomorphism $f:\mathcal V_\text{min}\to \mathcal V$ will be defined by the recursive formula (where $\bar \Theta$ is $\Theta$ without its quadratic part)
\be
\label{minimalmodelrecursion}
f^\star(z^a)=\zeta^a -G^{ab}f^\star\left(\partial_b\bar\Theta\right)=\zeta^a -G^a_bf^\star(Q^b-Q_0^b)\qquad (G^{ab}\equiv G^a_c \kappa^{bc})
\ee
Since $\partial\bar\Theta$ is $O(z^2)$, the right-hand side is  defined as a formal power series in $\zeta$ (cf.~\eqref{morphism}), and $f$ is an isomorphism in $K$-cohomology.

The sign in \eqref{minimalmodelrecursion} is fixed by demanding \eqref{minimalmodeleq1} (important in the sequel):
\begin{align}
C^a_b G^b_c Q^c=C^a_b G^b_c C^c_d z^d + C^a_b G^{bc}\partial_c\bar\Theta= C^a_b z^b + C^a_b G^{bc}\partial_c\bar\Theta\\ \label{minimalmodeleq1}
\implies f^\star\left(C^a_b G^b_c Q^c\right)=0\,.
\end{align}
The last equality can be rewritten
\be
\label{minimalmodelextreme}
f^\star\big(G^a_b \partial_a\Theta\big)=0
\ee
so we are in a sense ``solving the equations of motion'' derived from $\Theta(z)$. Then
\be
\Theta_\text{min}(\zeta)\equiv f^\star\Theta(z)\,,\quad (\kappa_\text{min})_{ab}\equiv P_a^c \kappa_{cd}P^d_b\,,\quad Q_\text{min}^a\equiv\kappa^{ab}\frac{\partial \Theta_\text{min}}{\partial\zeta^b}
\label{minimalmodelhamiltonian}
\ee
defines a minimal cyclic $L_\infty$-algebra if the master equation $\big(\Theta_\text{min},\Theta_\text{min}\big)_\text{min}=0$ is satisfied. A short calculation using $ \partial\Theta_\text{min}/\partial\zeta^a=P_a^b f^\star(\partial \Theta/\partial z^b)$ confirms
\be
\big(\Theta_\text{min},\Theta_\text{min}\big)_\text{min}=\kappa^{ab} P_a^c P_b^df^\star(\partial_c \Theta \partial_d \Theta)=f^\star \big(\Theta,\Theta\big)=0
\ee
where we used \eqref{minimalmodelextreme} and \eqref{kappablockdiagonal} in deriving the penultimate equality.

We have yet to show that $f$ is a morphism of $L_\infty$-algebras $\mathcal V_\text{min}\to \mathcal V$. (This is not especially illuminating.) This is condition \eqref{morphismcondition} which equivalently reads
\be
\big(f^\star \Theta, f^\star z^a\big)_\text{min}= f^\star\big(\Theta,z^a\big)\,.
\ee
Since $f$ is \emph{not} invertible, it does not automatically give a morphism of (graded) Poisson brackets despite $f^\star \kappa=\kappa_\text{min}$. We now loosely follow Kajiura \cite{kajiura} and first replace $f$ with an invertible morphism $F:\mathcal V\to\mathcal V$, again recursively defined,
\be
F^\star z^a=\hat z^a - G^a_b (F^\star \bar Q^b)\,,\quad (F^{-1})^\star \hat z^a=z^a + G^a_b \bar Q^b\,,\qquad (\bar Q^a=Q^a-Q_0^a)
\ee
where $\hat z^a$ is a copy of $z^a$. $F$ preserves the symplectic form due to $\kappa|_{\im (G K)}=0$: $F^\star \kappa=(\kappa_{ab} d\hat z^a d\hat z^b)/2$. It is therefore a symplectomorphism, which \emph{is} a morphism of Poisson algebras, and thus both $F$ and $F^{-1}$ are $L_\infty$-isomorphisms.

The transformed $L_\infty$-algebra structure is given by $\hat Q=F^\star\circ Q \circ (F^{-1})^\star$. The point of this redefinition is that the new vector field, $\hat Q$, is tangent to the subspace $P$ of cohomology representatives except for its ``non-minimal piece'':
\begin{align}
\hat Q&=Q^a(F^\star z)\frac{\partial (F^{-1})^\star \hat z^b}{\partial z^a}\frac{\partial}{\partial \hat z^b} \\
&= F^\star(Q^b+ G^b_c Q^a\partial_a \bar Q^c)    \frac{\partial}{\partial \hat z^b}\\
&=F^\star(C^b_c z^c + \bar Q^b+ G^b_c Q^a\partial_a \bar Q^c)    \frac{\partial}{\partial \hat z^b}\\
&= (C^b_c \hat z^c)\frac{\partial}{\partial \hat z^b}+ F^\star (-C^b_c G^c_d  \bar Q^d+ \bar Q^b+ G^b_c Q^a\partial_a \bar Q^c)    \frac{\partial}{\partial \hat z^b}\\
&= (C^b_c \hat z^c)\frac{\partial}{\partial \hat z^b}+ F^\star \bar Q^c P^b_c    \frac{\partial}{\partial \hat z^b}\,. \label{decompositiontheorem}
\end{align}
(The Jacobi identities in the form $Q^a\partial_a Q^b=0$ were used for the last step.)

With this setup, we see that $\hat f:\mathcal V_\text{min}\to\mathcal V$ given by a straightforward projection to the cohomology, $\hat f^\star \hat z^a=P^a_b \hat z^b\equiv \zeta^a$, is an $L_\infty$-algebra morphism yielding a minimal model for $\hat Q$. Since $\hat f^\star\circ F^\star=f^\star$, $f^\star$ is an $L_\infty$-algebra morphism. This completes the proof.


\end{proof}

The recursion \eqref{minimalmodelrecursion} clearly terminates for the purposes of determining $f$: a little counting shows that the $n$-th coefficient $f^{\phantom{n}a}_{n\,b_1b_2\dots b_n}$ \eqref{morphism} only depends on $C^a_{b_1\dots b_m}$ for $m\leq n$ and $f^{\phantom{m}a}_{m\,b_1b_2\dots b_n}$ for $m\leq (n-1)$. It replaces the usual sum over trees in the proof of the minimal model theorem (e.g.~Theorem 10.3.9 in \cite{loday2012algebraic}). This is highly suggestive of the combinatorics of the QFT 1PI generating functional, cf.~also \cite{kajiura}.

\section{The $L_\infty$-algebra of correlators}
Now consider a $D$-dimensional quantum field theory on a spacetime with coordinates $x^\mu$, with generating functional $Z[J]$ of correlation functions
\be
\delta ^n Z/\delta J(x_1)\dots \delta J(x_n)|_{J=0}=\langle \phi(x_1)\phi(x_2)\dots \phi(x_n)\rangle\,.
\ee Actually, for a gauge theory we consider the related functional $Z[J,\starred \Phi]$ where $\starred \Phi$ is a source for BRST transformations: differentiating $Z[J,\starred \Phi]$ with respect to $\starred \Phi$ leads to Ward identities for BRST symmetry. If a classical action is available, $Z[J,\starred \Phi]$ has a path integral expression
\be
\label{zj}
Z[J,\starred \Phi]=\int \mathcal D\phi\;\exp\left( \frac{i}{\hbar}S[\phi,\starred \Phi+\delta\Psi/\delta \phi] +\int dx\; J(x) \phi(x)\right)
\ee
where $S[\phi,\starred \phi]$ is the BV master action with bare fields $\phi$ (including ghosts) and antifields $\starred \phi$ and $\Psi$ is a gauge-fixing fermion. $S=S_\text{cl}+O(\hbar)$ satisfies the \emph{quantum} master equation,
\be
\label{quantummasterequation}
-2i\hbar \Delta S + (S,S)=0\,,\quad \Delta S=(-1)^{\phi}\frac{\delta^2 S}{\delta\phi(x)\delta\starred\phi(x)}
\ee
and its lowest order in $\hbar$ part $S_\text{cl}$ solves the classical master equation. (The antibracket $(\--,\--)$ here is the usual expression $\big(\phi(x),\starred \phi(y)\big)=\delta(x-y)$. The above is the usual setup of the BV formalism for arbitrary gauge theories, as reviewed comprehensively in e.g.~\cite{Gomis:1994he,Jurco:2018sby} and succinctly in e.g.~\cite{Arvanitakis:2017tla}. We omitted Lorentz or other indices on $\phi$.)

As an example of the above (from \cite{Henneaux:1992ig}) we briefly mention the Maxwell theory of a gauge potential $a_\mu(x)$ with field strength $f_{\mu\nu}\equiv 2\partial_{[\mu}a_{\nu]}$ on 4-dimensional Minkowski space. Fields $\phi$ are $a_\mu$, a ghost $c$ of degree $+1$, and in the non-minimal sector (only needed for gauge fixing) the Nakanishi-Lautrup field $b$ of degree $0$ and an ``antighost'' $\bar c$ of degree $-1$. Their antifields $\starred \phi$ are respectively $\starred a^\mu, \starred c, \starred{\bar c},\starred b$ of degrees $-1,-2,0,-1$ respectively. The gauge-fixing fermion enforcing e.g.~Lorenz gauge ($\partial^\mu a_\mu=0$) is $\Psi=\int i \bar c \partial^\mu a_\mu$. Then
\be
S[\phi,\starred\phi]=\int d^4 x\; -\frac{1}{4}f_{\mu\nu}f^{\mu\nu} + \starred a^\mu \partial_\mu c + i \starred{\bar c} b
\ee
solves \eqref{quantummasterequation}. $Z[J,\starred \Phi]$ is a path integral with measure $\int \mathcal Da \mathcal Dc \mathcal Db \mathcal D\bar c$ and we find in the integrand
\be
S[\phi,\starred \Phi+\delta\Psi/\delta \phi]=\int d^4 x\; -\frac{1}{4}f_{\mu\nu}f^{\mu\nu} + (\starred A^\mu-i\partial^\mu \bar c) \partial_\mu c + (i \starred{\bar C}-\partial^\mu a_\mu) b\,.
\ee
We thus see that $\starred A^\mu,\starred{\bar C}$ are classical sources for the corresponding gauge fixed BRST transformations. (The other components of $\starred \Phi$ drop out in this simple example.)

The Legendre transform \emph{in the sense of formal power series} \cite{jackson2017robust} with respect to $J(x)$ of the connected generating functional $W=\log Z$ is the 1PI generating functional $\Gamma[\Phi,\starred \Phi]$: define $\Phi(x)$ as a functional of $J$ and $\starred \Phi$ by
\be
\label{Phidefinitionlegendre}
\Phi(x)=\frac{\delta W[J,\starred \Phi]}{\delta J(x)}\,,
\ee
and invert it so $J$ is expressed using $\Phi,\starred \Phi$. Then
\be
\label{legendre}
\Gamma[\Phi,\starred \Phi]=-i\hbar\left(W- \int J(x)\Phi(x) dx\right)\,,
\ee
and $\delta\Gamma/\delta\Phi(x)=i\hbar (-1)^J J(x)$. Using this and integration by parts in the path integral,
\be
\label{zinnjustinequation}
(\Gamma,\Gamma)=\left\langle \hbar^{-2}\left(-2i\hbar \Delta S + (S,S) \right)\right\rangle_{J[\Phi]}\equiv\int\mathcal D\phi \frac{1}{\hbar^2 Z[J,\starred\Phi]}e^{\int J\phi} \left(-2i\hbar \Delta S + (S,S) \right)\,,
\ee
where $J= J[\Phi]$, $\langle \cdots\rangle_J$ is an expectation value in the presence of the source $J$ (see appendix B) and we defined an antibracket $\big(\Phi(x),\starred \Phi(y)\big)=\delta(x-y)$. Therefore, if $S$ solves the quantum master equation \eqref{quantummasterequation}, $\Gamma$ solves the \emph{classical} master equation, and vice versa inside correlators, since $J$ is an arbitrary source (cf.~the Schwinger-Dyson equations).

In this context the classical master equation $(\Gamma,\Gamma)=0$ for the 1PI functional is known as the Zinn--Justin equation \cite{ZinnJustin:1974mc}. It was originally found in the context of Yang-Mills, but can in fact be used to remove the divergences of fairly arbitrary field theories, as was done in \cite{Anselmi:1994ry} (see also the review \cite{Barnich:2000me}). Obstructions to the Zinn--Justin equation can be shown to correspond to local gauge anomalies; its validity implies Slavnov-Taylor identities. If the 1PI functional $\Gamma$ arises from a path integral involving a BV quantum master action $S$ as above, this is immediate from \eqref{zinnjustinequation}. (We refer to the review \cite{Gomis:1994he} sections 8.4 and 8.5 for more on this point and for a proof of \eqref{zinnjustinequation}.) We prefer however to think of $\Gamma$ as the fundamental object, not necessarily expressed as an $\hbar$-expansion or path integral.


We take $\Gamma$ along with the antibracket pairing between $\Phi$ and $\starred \Phi$ to define the cyclic $L_\infty$-algebra of correlators. Assuming $(\Gamma,\Gamma)=0$ we only need check that
\be
\frac{\delta \Gamma}{\delta \Phi}=\frac{\delta \Gamma}{\delta \starred\Phi}=0 \qquad \text{at }\Phi=\starred \Phi=0
\ee
to match the form of \eqref{thetadef}; the structure constants of the $L_\infty$-algebra of correlators are then obtained by expanding $Q=(\Gamma,\--)$ in $(\Phi,\starred \Phi)$ around $\Phi=\starred \Phi=0$.
These conditions have physical interpretations. $\delta \Gamma/\delta \Phi=0$ at $\Phi=0$ (which we need anyway for the Legendre transform) says that $\phi(x)$ has vanishing vacuum expectation value. This is usually violated in the context of spontaneous symmetry breaking, but can be remedied by a constant field shift for QFTs with Poincar\'e-invariant vacuum on Minkowski space. The other condition, $\delta \Gamma/ \delta \starred \Phi=0$, expresses the absence of BRST anomalies.

This setup applies of course to the closed bosonic string field theory of Zwiebach \cite{Zwiebach:1992ie}. The ``quantum string action'' therein ($S$ in our notation) is written as a genus expansion --- equivalently, an $\hbar$-expansion --- and satisfies the quantum master equation \eqref{quantummasterequation} as a consequence of the identities satisfied by the string products; in other words, as a consequence of the ``loop $L_\infty$-algebra'' \cite{Markl:1997bj} Jacobi identities (which at tree-level reduce to the well-known Jacobi identities of the \emph{ordinary} $L_\infty$-algebra of closed string field theory). We have here translated these identities into the Zinn--Justin equation for the string field theory 1PI effective action $\Gamma$. The upshot is that the ordinary $L_\infty$-algebra structure defined by $\Gamma$ encodes both the well-known $L_\infty$-algebra of tree-level closed string field theory as well as the higher-genus contributions that deform it into a loop $L_\infty$-algebra; the price to be paid is that the $L_\infty$-algebra of $\Gamma$ is not defined over $\mathbb C$, but rather over $\mathbb C[\hbar]$ (complex formal power series in $\hbar$).

Before moving on we acknowledge that we have not actually specified the underlying vector space $\mathcal V$ on which the $L_\infty$ brackets act, so the construction in this section is so far formal. Fixing $\mathcal V$ at this level of generality is difficult. For the more concrete case of a scalar QFT on Minkowski space, we define $\mathcal V$ in the next section; it is simply a direct sum of on-shell wavefunctions (i.e.~states appearing in the S-matrix) and Schwartz functions (i.e.~functions of rapid decrease). One expects a similar picture for more general theories.

\section{The $L_\infty$-algebra of the S-matrix}
Our claim here is that the LSZ reduction formula can be interpreted as the quasi-isomorphism appearing in the above proof of the minimal model theorem for cyclic $L_\infty$-algebras. To this end we recall and clarify the observation originally due to Jevicki and Lee \cite{Jevicki:1987ax} (see also \cite{Fukuda:1988ka,Kim:1996nd} and the textbooks \cite{Nair:2005iw,siegel1999fields}) that the S-matrix functional \cite{Arefeva:1974jv} is obtained from the 1PI functional $\Gamma[\Phi,\starred \Phi]$ by extremising it.

We assume here the same setup as in the scalar LSZ  formula \eqref{lsz}. Our QFT only has one real scalar field operator $\phi(x)$ with 2-point function (propagator)
\be
\label{KLspectral}
G(x-y)\equiv\langle \phi(x)\phi(y)\rangle=\int_0^\infty d(\mu^2)\; \rho(\mu^2) G_F(x-y;\mu^2)
\ee
where $G_F(x;\mu^2)$ is the usual Feynman propagator for the mass $\mu^2$ Klein-Gordon equation, satisfying $(\square +\mu^2)G_F(x;
\mu^2)=-i\hbar\delta(x)$. This is the K\"all\'en--Lehmann spectral representation \cite{Kallen:1952zz,Lehmann:1954xi}. The spectral function $\rho(\mu^2)$ takes the form
\be
\label{spectralfunction}
\rho(\mu^2)= Z\delta(m^2-\mu^2)+ \sigma(\mu^2)\,,\quad 0<Z<1
\ee
appropriate for a scalar QFT whose asymptotic free particle states have mass $m^2$, so the smooth function $\sigma(\mu^2)$ is only non-zero above a threshold $m_\text{thresh}^2>m^2$ for production of multiparticle states. We assume furthermore that there are no asymptotic states besides the ones created by $\phi(x)$, and suggest a remedy in the Discussion.

These assumptions are fairly restrictive, but our arguments should apply \emph{mutatis mutandis} to any theory where states entering the S-matrix are all massive. A considerable technical simplification for scalar theories is that we need not introduce gauge symmetry, so we can without loss of generality let $\Gamma$ and $W$ be independent of $\starred \Phi$. Therefore, the Zinn--Justin equation $(\Gamma,\Gamma)=0$ is trivially satisfied.



The LSZ formula \eqref{lsz} relates the connected S-matrix generating functional $\mathcal A[\varphi]$ to $W[J]$ by evaluation of the latter on the source $J_\varphi$, defined by
\be
\label{jphidef}
\int dx\; J_\varphi(x) f(x)= \frac{i}{\sqrt{Z}}\int dx\;\varphi(x) (\square_x+m^2) f(x)
\ee
for functions $f$ such that the right-hand side makes sense.
Here $\varphi(x)$ is the wavefunction of an asymptotic 1-particle state, i.e.~an S-matrix state:
\be
\label{kleingordon}
(\square_x+m^2)\varphi(x)\equiv(\partial_t^2 -\nabla^2 +m^2)\varphi(x)=0\,,
\ee
so the integrand above is in fact a total derivative, and $J_\varphi$ is a somewhat bizarre distribution supported at infinity (more on this later). A subtlety here  is that $\mathcal A[\varphi]$ has a non-zero universal quadratic part associated to  trivial $1 \to 1$ scattering, while the quadratic in $\varphi$ part of $W[J_\varphi]$ vanishes for our gapped scalar QFT using \eqref{KLspectral} and \eqref{spectralfunction} (see \cite{Itzykson:1980rh} section 5-1-5). Thus, $W[J_\varphi]$ is the generating functional of \emph{non-trivial} connected S-matrix elements, and we will write $\mathcal A[\varphi]=W[J_\varphi]$, implicitly discarding the quadratic part.



To rewrite this in terms of the 1PI functional, we need to solve $\delta \Gamma/\delta \Phi(x)=i \hbar J(x)$ for $\Phi$ in terms of $J$. This is the (inverse) Legendre transform of formal power series \cite{jackson2017robust}. Since $G(x)$ of \eqref{KLspectral} is the inverse of $\delta^2\Gamma/\delta\Phi^2|_{\Phi=0}$ up to prefactors, if $\bar\Gamma$ is $\Gamma$ without its quadratic part, we find the general solution
\be
\Phi(x)= \int dy\; G(x-y)\left(J(y)+\frac{i}{\hbar}\frac{\delta \bar\Gamma}{\delta \Phi(y)}\right)\,.
\ee

For $J= J_\varphi$, definition \eqref{jphidef} and \eqref{KLspectral}, \eqref{spectralfunction}, \eqref{kleingordon} give
\be
\label{GJvarphi}
\int dy\; G(x-y) J_\varphi(y)=\hbar\sqrt{Z} \varphi(x)\,;
\ee
here crucially the multiparticle contribution $\sigma(\mu^2)$ to the spectral function \eqref{spectralfunction} drops out, effectively replacing $G(x)$ by the (free) propagator $G_F(x\,;m^2)$ up to a factor of $Z$. If $J_\varphi$ is interpreted as a ``source at infinity'' for an incoming/outgoing on-shell state with wavefunction $\varphi(x)$, this calculation states that the multiparticle states fail to contribute to the time evolution of $\varphi$ from $x^0=\pm \infty$ to finite values. The mass gap $m^2_\text{thresh}>m^2$ of \eqref{spectralfunction} is crucial to this calculation.

We therefore find the following recursive formula defining $\Phi$ as a formal power series in the wavefunction $\varphi$ (cf.~\eqref{minimalmodelrecursion})
\be
\label{Phirecursion}
\Phi_\varphi(x)=\hbar \sqrt{Z} \varphi(x)+\frac{i}{\hbar}\int dy\; G(x-y) \frac{\delta \bar\Gamma}{\delta \Phi(y)}
\ee
which leads to the following formula for the S-matrix functional
\be
\mathcal A[\varphi]=i\hbar^{-1} \Gamma[\Phi_\varphi]+\int J_\varphi(x)\Phi_\varphi(x) dx\,.
\ee

So far we have never dropped any boundary terms. We would like to now drop the second term  in $\mathcal A[\varphi]$ above, which is proportional to the total derivative
\be
\label{problemterm}
\int dx\; \varphi(x) (\square +m^2)\Phi_\varphi(x)=\int dx\; \partial_\mu\big( \Phi_\varphi(x) \partial^\mu \varphi(x)- \varphi(x) \partial^\mu \Phi_\varphi(x) \big)
\ee
whenever $\varphi(x)$ solves the Klein-Gordon equation \eqref{kleingordon}. This is anyway zero to order $O(\varphi^2)$. Obviously we cannot in general prove this vanishes (e.g. in the sense of formal power series) without estimates on 1PI correlators and without specifying the space of $\varphi(x)$. For $\varphi(x)$ an appropriate space is the space $V_{\rm rwp}$ of \emph{regular wave packets}, i.e. of smooth solutions to the Klein-Gordon equation \eqref{kleingordon} with initial data of compact support in momentum space:
\be
\label{regularwavepacket}
\varphi(x)=\int \frac{d^{D-1}k}{  \sqrt{(2\pi)^{D-1} 2 E_{\vec{k}} }   } \;\big(\alpha(\vec k) e^{i(E_{\vec{k}}t + \vec{k}\cdot \vec{x})} +\bar\alpha(\vec k) e^{-i(E_{\vec{k}}t + \vec{k}\cdot \vec{x})}\big)\,,\quad E_{\vec{k}}=\sqrt{|\vec{k}|^2+m^2}
\ee
(where $\alpha({\vec{k}})$ is basically the spatial Fourier transform of the initial data $\varphi|_{x^0=t=0}$ for $\varphi(x)$, $\bar\alpha(\vec k) $ is its complex conjugate, and both are of compact support by assumption. Regular wave packets can approximate momentum eigenstates arbitrarily well, which is the primary consideration here.) These always vanish at infinity  (Theorem XI.17 of \cite{reed1979methods}), so assuming that $\Phi_\varphi(x)$ does not blow up at infinity, \eqref{problemterm} should not contribute to the S-matrix. This is also argued to be the case in \cite{Jevicki:1987ax,Kim:1996nd}. As a final justification, we note that with the choice of $\mathcal V$ specified later in this section it is clear this total derivative term vanishes.

The upshot is the following formula for the S-matrix functional involving the 1PI functional alone \cite{Jevicki:1987ax} (also \cite{Fukuda:1988ka,Kim:1996nd}):
\be
\label{smatrixfunctionaljevicki}
\mathcal A[\varphi]=\frac{i}{\hbar} \Gamma[\Phi_\varphi]\,.
\ee
One interpretation is that one obtains the S-matrix by evaluating $\Gamma[\Phi]$ on the solution of $\delta \Gamma/\delta \Phi(x)=0$: writing
\be
\Gamma[\Phi]=\frac{1}{2}(i\hbar)\int dx\int dy \; K(x-y)\Phi(x)\Phi(y)+ O(\Phi^3)\,,
\ee
we find (the $i\hbar$ right above absorbs superfluous factors below)
\be
\label{propK}
\int dz\; K(x-z) G(z-y)= \delta(x-y)\,.
\ee
Using this, \eqref{GJvarphi} yields
\be
\label{Kvarphi}
\int dy\; K(x-y)\varphi(y)=(\hbar \sqrt{Z})^{-1} J_\varphi(x)\,.
\ee
The right-hand side vanishes when integrated against any $f(x)$ of compact support by \eqref{jphidef}, so in some sense $\varphi(x)$, which is annihilated by the Klein-Gordon operator \eqref{kleingordon}, is also annihilated by the operator $K$ derived from the 1PI effective action. (This is a check that $\varphi$ satisfies the Klein-Gordon equation with renormalised mass, as it anyway must if it is the wavefunction of a scattering state.) Of course, the right-hand side of \eqref{Kvarphi} is very much non-zero. This apparent tension is resolved by interpreting $J_\varphi(x)$ as a change of boundary conditions: the equation $K f(x)=J_\varphi(x)+ J(x)$ for $f(x)$ is interpreted as $K(g(x))=J(x)$ with $g(x)=f(x)-\hbar \sqrt{Z}\varphi(x)$ lying in $V_\text{Schwartz}$ i.e.~the space of functions of rapid decrease; the point here being that $K$ is invertible when restricted to $V_\text{Schwartz}$\footnote{This is easiest to see from the the K\"all\'en--Lehmann expression for the exact propagator $G$ \eqref{KLspectral} and \eqref{spectralfunction}, which should really be interpreted in momentum space, where $G$ is a multiplication operator. Assuming $G$ exists in this context as a multiplication operator on $V_\text{Schwartz}$, doing a Wick rotation shows it has an inverse, which is $K$ by definition. The position space expressions used here are then obtained by Fourier transform, which is anyway an isomorphism on $V_\text{Schwartz}$. We refer to \cite{Zeidler:2006rw} chapter 14 for the relevant analysis background.}. Analogously, we see that $\Phi_\varphi(x)$ of \eqref{Phirecursion} is the unique solution of
\be
\frac{\delta \Gamma}{\delta \Phi(x)}=i\hbar \int dy\; \big(K(x-y)\Phi(y)\big) + \frac{\delta \bar\Gamma}{\delta \Phi(x)}=0
\ee
if we assume $\Phi(x)-\hbar \sqrt{Z}\in V_\text{Schwartz}$. At tree level, this reduces to the well-known recipe for tree-level S-matrix elements from the classical action, recently exploited in e.g. \cite{Adamo:2018srx,Adamo:2017nia}.

The argument above is clarified by considering the free theory: in that case $Z=1$, $K$ is proportional to $(\square + m^2)$, and $G$ is the \emph{Feynman} propagator. The choice of Feynman boundary conditions for the propagator is related to the choice of boundary conditions in the path integral, which are in turn fixed by our decision to calculate \emph{vacuum} correlators.

We now articulate the $L_\infty$-algebraic interpretation. We define the hamiltonian $\Theta[\Phi]$ and antibracket defining the $L_\infty$-algebra of correlators as
\be
\label{scalarLinfinityalgebracorr}
\Theta[\Phi]=-\frac{i}{\hbar}\Gamma[\Phi]\,,\quad \big(\Phi(x),\starred\Phi(y)\big)=\delta(x-y)\,, \quad \deg \Phi(x)=0\,,\quad \deg \starred \Phi(x)=-1\,.
\ee
$\Phi(x)$ and $\starred \Phi(x)$ play the role $z^a$ did in the general discussion previously. In particular $\Phi(x)$, $\starred \Phi(x)$ are \emph{not} the values of fixed functions at some spacetime point $x$, but rather they are linear functionals mapping functions $f:\mathbb R^{D}\to \mathbb R$ to their values at a spacetime point $x$ (much like $z^a\in \mathcal V^\star$ is a linear map $\mathcal V\to\mathbb R$.)\footnote{This is apparently called the ``fundamental confusion of calculus'' and is elaborated on in \cite{Jurco:2018sby} in this context. It is perhaps clearer to explain this way: since $z^a$ are a basis of $\mathcal V^\star$, if $v=v^a T_a$ with $v^a\in\mathbb R$ and $T_a\in \mathcal V$ are a basis, the map $\mathcal V\to\mathbb R$ defined by $v\to v^a$ (for any specific choice of index $a$) is simply $\langle z^a|v\rangle= \langle z^a|T_b\rangle v^b=\delta^a_b v^b=v^a$. The confusion is that $\deg v^a=0$ (since $v^a\in\mathbb R$ is a number) but $\deg z^a=-\deg T_a$ (which is anyway necessary for $\langle z^a|T_b\rangle=\delta^a_b$ since numbers are in degree zero). With regard to determining the brackets on $\mathcal V$ from $Q$ acting on polynomials in $z^a$, we have e.g. $Qz^a= C^a_b z^b+O(z^2)$ (corresponds to \eqref{bracketsfromTheta}), $Kv= KT_b v^b= (C_b^a v^b)T_a \implies v^a\to C^a_b v^b$ (corresponds to \eqref{unarybracketscalarfieldtheory}).} The various $L_\infty$-algebra brackets act on the space $\mathcal V$ spanned by their duals. Since $\deg \Phi=0\,,\deg \starred \Phi=-1$, $\mathcal V$ is concentrated in degrees $0$ and $+1$ respectively. At degree 0 we have ordinary scalar fields $\phi(x)$ while at degree 1 we have degree-shifted scalars we will write $c \phi(x)$  using a degree $+1$ formal constant $c$.

Since $Q=\int dx\;\big(\Theta,\starred \Phi(x)\big)\,\delta/\delta\starred\Phi(x)$, the $L_\infty$-algebra structure constants can be read off the expansion in $\Phi(x)$ of
\be
\label{bracketsfromTheta}
\big(\Theta,\starred \Phi(x)\big)=\int dy\; K(x-y)\Phi(y) + O(\Phi^2)\,.
\ee
In particular the unary bracket $K$ reads
\be
\label{unarybracketscalarfieldtheory}
K \phi(x)=c\int dy\; K(x-y) \phi(y)\,,\quad K(c\phi(x))=0\,,
\ee
and the 1PI $n$-point correlators $n\geq 3$ similarly define the $(n-1)$-ary bracket $[\phi_1,\phi_2,\dots \phi_n](x)$. (Any bracket any of whose arguments involves $c$ vanishes. All the Jacobi identities are thereby satisfied.)

Therefore the vector space of the $L_\infty$-algebra of correlators is
\be
\mathcal V=V\oplus V[+1]\,,\qquad V\xrightarrow{K} V[+1]
\ee
where $V$ is the following space of scalar fields $\phi(x)$
\be
\label{spaceofscalars}
\phi(x)=\varphi(x)+\phi_{\rm S}(x)\,,\quad \text{i.e. } V=V_{\rm rwp}\oplus V_\text{Schwartz}
\ee
a direct sum of regular wave packets $\varphi(x)$ (i.e.~wavefunctions of asymptotic 1-particle states) along with $\phi_{\rm S}(x)\in V_\text{Schwartz}$ lying in the Schwartz space $V_\text{Schwartz}$ of rapidly-decreasing functions.

With this choice, the cohomology of $K$ at degree 0 is  the regular wave packets:
\be
K\phi(x)=0 \iff \int dy \; K(x-y)  \big(\varphi(y)+\phi_{\rm S}(y)\big)=0\iff \phi_{\rm S}(x)=0\,.
\ee
(There is no tension with \eqref{Kvarphi} here since $J_\varphi(x)$ is not an element of $\mathcal V$; it has been projected out. Hopefully our use of $K$ in both places to denote slightly different operators is not confusing.)
Since $\phi_{\rm S}(x)=K G\phi_{\rm S}(x)$ in slightly abusive notation it is clear that a space of cohomology representatives at degree $1$ is again the space of regular wave packets. Therefore, we find the underlying vector space of the putative minimal $L_\infty$-algebra
\be
\mathcal V_\text{min}=V_\text{rwp}\oplus V_\text{rwp}[+1]\,.
\ee

We have been coy about discussing the cyclic inner product on the $L_\infty$-algebra of correlators (or that of the S-matrix) so far. The reason is that our sins --- in recklessly taking duals of infinite-dimensional spaces --- have now caught up with us. The canonical antibracket $\big(\Phi(x),\starred \Phi(y)\big)$ of \eqref{scalarLinfinityalgebracorr} formally defines the degree -1 symplectic form $\kappa=\int dx\; \delta\Phi(x)\delta \starred \Phi(x)$ which is a pairing $V\times V[+1]\to \mathbb R$. A near-identical construction appears in Costello's treatment of $\phi^4$ theory (Example 5.1 of \cite{Costello:2007ei}), the difference being that he works on a compact Euclidean spacetime, where this $\kappa$ is well-defined.

We however must work on Minkowski spacetime in order to treat the on-shell external states. The above ``na\"ive'' bilinear form $\kappa$ blows up for regular wave packets $\varphi(x)$: indeed from \eqref{regularwavepacket} we calculate
\be
\int dx\; \varphi_1(x) \varphi_2(x)=\int dt\;\int \frac{d^{D-1}k}{E_{\vec{k}}} {\rm Re}\:[\alpha_1(\vec{k})\bar \alpha_2(\vec{k}) + \alpha_1(\vec{k})\alpha_2(\vec{-k}) e^{i(2 E_{\vec{k}})t}]\,.
\ee
The first term blows up due to a factor $\int dt \; 1$, while the second is singular due to the oscillatory integral $\int dt \; \exp(i2 E_{\vec{k}})t\propto \delta(2 E_{\vec{k}})$ which should in some sense vanish since $E_{\vec{k}}$ is bounded away from zero by $m>0$. This requires regularisation. We thus smear $\varphi(x)$ around the mass-shell in momentum space by replacing
\be
\varphi(x)=\int dk_0 \frac{d^{D-1}k}{  \sqrt{(2\pi)^{D-1} 2 E_{\vec{k}}} } \; \delta(k_0- E_{\vec{k}})\big(\alpha(\vec k) e^{ikx} +\bar\alpha(\vec{k}) e^{-ikx} \big)
\ee
with
\be
\varphi_\varepsilon(x)=\int dk_0 \frac{d^{D-1}k}{  \sqrt{(2\pi)^{D-1} 2 E_{\vec{k}}} } \; \delta_\varepsilon(k_0- E_{\vec{k}})\big(\alpha(\vec k) e^{ikx} +\bar\alpha(\vec{k}) e^{-ikx} \big)
\ee
where $\delta_\varepsilon(k_0)$ is an e.g. Gaussian approximation of the Dirac delta:
\be
\delta_\varepsilon(k_0)=\frac{1}{2\pi} \int dt\; \exp( i k_0 t -\varepsilon t^2)=\frac{1}{\sqrt{4\pi\varepsilon}}e^{-(k_0)^2/4\varepsilon}\,,\qquad \varepsilon >0\,.
\ee
This can be seen as an $i\varepsilon$-prescription. Then $\int dx\; \varphi_{1,\varepsilon}(x) \varphi_{2,\varepsilon}(x)$ reads
\be
\sqrt{\frac{\pi}{2}}\int \frac{d^{D-1}k}{E_{\vec{k}}} {\rm Re}[\alpha_1(\vec{k})\bar \alpha_2(\vec{k}) \varepsilon^{-1/2}+\alpha_1(\vec{k})\alpha_2(\vec{-k})\varepsilon^{-1/2}e^{- E_{\vec{k}}^2/(2\varepsilon)}]
\ee
where the first term diverges like $\varepsilon^{-1/2}$ as $\varepsilon\to 0$, but the second term vanishes.

We therefore define a regularised cyclic inner product $\kappa$ in the $L_\infty$-algebra of correlators by rescaling $\varphi(x)$ in \eqref{spaceofscalars} by $(2\varepsilon/\pi)^{1/4}$:
\begin{align}
\label{cyclicinnerproductscalarQFT}
\kappa(\phi_1,c\phi_2)&=\lim_{\varepsilon\to 0}\int dx\; \left( \left(\frac{2\varepsilon}{\pi}\right)^{1/4} \varphi_{1,\varepsilon}(x) + \phi_{\rm 1,S}(x) \right) \left( \left(\frac{2\varepsilon}{\pi}\right)^{1/4} \varphi_{2,\varepsilon}(x) + \phi_{\rm 2,S}(x) \right)\\
&=\int \frac{d^{D-1}k}{E_{\vec{k}}}\; {\rm Re}[\alpha_1(\vec{k})\bar \alpha_2(\vec{k})] + \int dx\; \phi_{\rm 1,S}(x) \phi_{\rm 2,S}(x)\,.
\end{align}
(Recall $\mathcal V=V\oplus V[1]$ where we write $V[1]$ with the degree $+1$ formal constant $c$.)
Not only did we absorb the divergence --- yielding a positive-definite inner product on $V_{\rm rwp}$ --- but we also rendered $V_{\rm rwp}$ $\kappa$-orthogonal to the Schwartz functions $\phi_{\rm S}(x)$. This realises a Hodge-Kodaira decomposition
\be
\mathcal V=P\oplus P^\perp\,,\qquad   P=\mathcal V_\text{min}= V_{\rm rwp}\oplus V_{\rm rwp}[+1]\,,\quad  P^\perp=V_\text{Schwartz}\oplus V_\text{Schwartz}[+1]
\ee
in the sense of the proof of the minimal model theorem in the preceding section, for which the degree -1 map $G:\mathcal V\to \mathcal V$ defined by (where again $G(x-y)$ is \eqref{KLspectral})
\be
G:V[+1]\to V\,,\qquad G(c \varphi(x))=0\,,\quad G(c\phi_{\rm S}(x))=\int dy\; G(y-x) \phi_{\rm S}(y)
\ee
and vanishing otherwise, is a ``propagator'' in that same sense.

We are not quite done yet however, as there are outstanding analytic issues in the presence of interactions (i.e.~non-vanishing binary and higher \lf{}-brackets), as was originally pointed out in \cite{Macrelli:2019afx}. These even appear at tree level, where $\Gamma$ is the classical action: in the presence of an e.g.~$\phi^3$ interaction, $\Gamma$ fails to be defined because the integral $\int dx\; \phi^3(x)$ blows up when $\phi\in V_{\rm rwp}$. This is remedied in \cite{Macrelli:2019afx} by (effectively) inserting a position-dependent Gaussian in the $\phi^3$ term, at the price of Lorentz invariance. In the general case, beyond tree level, we regard these issues as part of the definition of $\Gamma$.

\bigskip

The upshot is that we have now realised the setting of the proof of the minimal model theorem as given in the preceding section for cyclic $L_\infty$-algebras. Crucially, the recursive definition of the minimal model brackets via \eqref{minimalmodelrecursion} is identical to the recursive formula \eqref{Phirecursion} for $\Phi(x)$ originally derived by the usual Legendre transform of formal power series (relating the 1PI functional $\Gamma$ to the generating functional $W$ of connected correlators), as applied to the asymptotic source $J_\varphi(x)$ sourcing incoming/outgoing 1-particle states. The cyclic $L_\infty$-algebra of the S-matrix this procedure yields is defined by the hamitonian $\Theta[\varphi]$ which is the S-matrix generating functional $\mathcal A[\varphi]$ \eqref{smatrixfunctionaljevicki} up to a sign, with cyclic inner product the restriction of \eqref{cyclicinnerproductscalarQFT} to the space of asymptotic 1-particle states.

\section{Discussion}
We argued that the most basic objects one usually cares about in quantum field theory --- the S-matrix, and vacuum correlators --- have a not-widely-appreciated $L_\infty$-algebraic structure. Moreover, we claim this structure is \emph{natural}:
\begin{itemize}
	\item the $L_\infty$ Jacobi identities are the \emph{non-anomalous} Slavnov-Taylor identities;
	\item the $L_\infty$-algebra of the S-matrix is obtained from the $L_\infty$-algebra of correlators by a canonical construction, that is the \emph{minimal model theorem};
	\item Maurer-Cartan elements \eqref{MaurerCartan} are the extrema of the 1PI functional $\Gamma$, corresponding to the moduli space of vacua \`a la Coleman-Weinberg \cite{Coleman:1973jx}.
\end{itemize}
The facts involving the $L_\infty$-algebra of correlators alone follow by trivial observations from Zinn--Justin's 1974 work \cite{ZinnJustin:1974mc}; we have only given a dictionary that translates from his ``antifield'' or BV language to $L_\infty$ language. However, the relation to the S-matrix (outside of string field theory \cite{Witten:1992yj,Verlinde:1992qa,Kajiura:2001ng,Munster:2012gy,Konopka:2015tta} and the more recent independently-derived results of \cite{Nutzi:2018vkl} for tree-level gravity and Yang-Mills) is new.

An orthogonal but not incompatible interpretation of our results is that $L_\infty$-algebraic structures do not characterise closed string field theory (where they were originally discovered \cite{Zwiebach:1992ie}), but are instead immanent to quantum field theory \emph{with a choice of vacuum}. Supporting this is the last claim above: in the context of the philosophy which associates an $L_\infty$-algebra to a deformation problem \cite{schlessinger2012deformation}, we see from the Coleman-Weinberg \cite{Coleman:1973jx} calculation that the Maurer-Cartan moduli space \eqref{MaurerCartan} of the $L_\infty$-algebra of correlators corresponds to the vacuum moduli space.

The validity of our results hinges largely on
\begin{enumerate}
	\item having a well-defined 1PI effective action $\Gamma$, known either exactly, or to some order in $\hbar$, satisfying the Zinn-Justin equation $\big(\Gamma,\Gamma\big)=0$;
	\item (for the Minkowski space S-matrix:) having a mass gap, so the LSZ formula  and K\"all\'en--Lehmann spectral representation \cite{Kallen:1952zz,Lehmann:1954xi} are valid;
	\item (also for the S-matrix:) there being no bound states.
\end{enumerate}

Requirement 3.~could conceivably be relaxed in general: if a bound state appearing in the S-matrix is created by a composite operator $\mathcal O[\phi](x)$ (where $x$ here is some sort of collective coordinate), one could source it using a current $J_\mathcal O(x)$ in the path integral to arrive at a generating functional $Z[J,J_\mathcal O,\starred \Phi]$, then take a logarithm and do a Legendre transform on both $J$ and $J_{\mathcal O}$ to try to derive a Zinn--Justin equation for a new kind of 1PI functional. This direction has been pursued in \cite{Anselmi:2012qy}, including a Zinn--Justin equation for what is called a ``master functional'' therein.

With regard to requirement 1.~we emphasise that it is not necessary that $\Gamma$ be known exactly; our arguments work if for instance we have an expansion
\be
\Gamma=\Gamma_0 +\hbar \Gamma_1 +\dots+\hbar^n \Gamma_n +O(\hbar^{n+1})
\ee
and we do not care to calculate the higher corrections. Here we only need replace the field $\mathbb R$ or $\mathbb C$ of scalars of the $L_\infty$ algebra with the ring $\mathbb R[\hbar]/\{\hbar^{n+1}\}$ (where we quotiented formal power series in $\hbar$ by the ideal generated by $\hbar^{n+1}=0$). This produces a large class of examples, the most accessible of which are tree-level theories, where $\Gamma$ reduces to the classical action by the usual stationary-phase argument.

Relatedly, we have not considered questions of renormalisation. We simply assume $\Gamma$ has been defined at some mass scale $\mu^2$ so condition 1.~is satisfied. This does lead to \emph{a priori} inequivalent $L_\infty$-algebras for each $\mu^2$. We prefer to resolve this in the future. One might expect to make contact with the recent treatment of renormalisation in the BV formalism \cite{costello2011renormalization} which is however based on the quantum master action $S$, not the 1PI functional $\Gamma$ (linked by \eqref{zj}, \eqref{legendre}, and \eqref{zinnjustinequation}).

As a mathematical aside we mention that the above relation between $S$ and $\Gamma$ appears to associate to any loop $L_\infty$-algebra \cite{Markl:1997bj} (say over $\mathbb R$) an ordinary $L_\infty$-algebra over $\mathbb R[\hbar]$ through a ``perturbative path integral'' expression. The Jacobi identities on either side are then related by the obvious generalisation of \eqref{zinnjustinequation}. The calculations showing the correspondence between the respective cohomologies (see e.g. \cite{Gomis:1994he} sections 8.4, 8.5) should be rigorous given that perturbative path integrals can be formalised, as in e.g. \cite{braun2018minimal}. This appears to lead to a compact reformulation of the notion of loop $L_\infty$-algebra.

Requirement 2.~--- relevant for the S-matrix on Minkowski space --- is the hardest to satisfy. The LSZ formula and the K\"all\'en--Lehmann spectral representation fail in the more interesting context of gauge theories: the former because gauge theories usually have massless degrees of freedom (i.e.~no mass gap) which contradicts the assumption of free asymptotic 1-particle states, and the latter because the space of states is not a Hilbert space when ghosts are present (they have negative norm). Of these, the failure of the LSZ formula is the more serious one. Another way to phrase it is that infrared divergences obstruct the construction of an S-matrix. Of course, at tree-level there are no divergences and the S-matrix can indeed be obtained according to \eqref{smatrixfunctionaljevicki}.

\bigskip
We now suggest applications and generalisations. One immediate application is to \emph{tree-level} theories, which already generalises the aforementioned known results. In the paper \cite{Macrelli:2019afx} (which appeared at the same time as version 1 of this paper) this is pursued with the remarkable result that the minimal model theorem leads to practical recursion relations for scattering amplitudes for any tree-level theory. In our language this can be understood as follows: from \eqref{minimalmodelhamiltonian} we see that an $n$-point amplitude is built off the quantities $f^{\phantom{n}a}_{n\,b_1b_2\dots b_n}$ calculated from \eqref{minimalmodelrecursion}. These depend only on the $m\leq (n-1)$-point amplitudes as well as the $n$-ary and \emph{lower} structure constants of the \lf{}-algebra of correlators; at tree level, the latter are essentially the structure constants of the original lagrangian and are therefore known. In fact the $f^{\phantom{n}a}_{n\,b_1b_2\dots b_n}$ are almost the same as the Berends-Giele currents \cite{Berends:1987me} as might be guessed from the fact they have the same index structure, involving $n$ on-shell legs and a single off-shell leg. The recursion \eqref{minimalmodelrecursion} is then effectively identical to Berends-Giele recursion. While the recursion \eqref{minimalmodelrecursion} also exists at loop level, it is less practical because the \lf{}-structure constants $C^a_{b_1\dots b_n}$ it involves are then the correlators of the theory and are not known a priori.

Consider furthermore the problem of \emph{S-matrix equivalence} i.e.~of finding which field redefinitions lead to equivalent S-matrices \cite{Chisholm:1961tha,Kamefuchi:1961sb,Kallosh:1972ap}. Clearly the necessary and sufficient condition is that a field redefinition induces a quasi-isomorphism of the corresponding $L_\infty$-algebras of 1PI correlators. Checking this is not entirely straightforward, but probably easier than recalculating the S-matrix. For linear field redefinitions this is easy, however: consider e.g.~acting with a Poincar\'e transformation on $\phi(x)$ of the last section. Since the result is linear in $\phi(x)$ we find $\Phi(x)$ transforms the same way (given that the vacuum is Poincar\'e invariant). This is trivially an $L_\infty$-isomorphism of cyclic $L_\infty$-algebras, showing Poincar\'e invariance of the S-matrix.

The $L_\infty$-algebraic perspective is well-suited to problems involving determining deformations of QFTs (or, rather, their correlation functions). Given our results, we see that the deformation problem is controlled in general by cyclic $L_\infty$-algebra cohomology \cite{Penkava:1994mu}, which can be reduced to the cohomology of the cyclic minimal model (as is obvious from \eqref{decompositiontheorem}), i.e.~the S-matrix. In fact applications vaguely along those lines have already appeared in the form of \cite{Blumenhagen:2018kwq,Blumenhagen:2018shf} wherein non-commutative deformations of Chern-Simons and Yang-Mills were determined through an ``$L_\infty$-bootstrap''.

Finally, consider the problem of extending our results to conformal field theories. As explained above, the construction in this paper does not directly apply\footnote{As there is no mass gap or obvious notion of free asymptotic states to scatter.} so it is unclear what a ``minimal model'' of a CFT in the sense of the minimal model theorem should correspond to (again, this is a different notion than that of a minimal model CFT). A related curiocity is that all 2-point functions of any CFT are fixed by conformal symmetry. Something similar happens to the S-matrix functional: $1\to 1$ scattering is trivial, so the quadratic term in any S-matrix functional is universal, as we already pointed out in the previous section. This suggests we could discard the 2-point functions in a CFT (since there is effectively no information therein) and see if $n$-point CFT correlators for $n\geq 3$ are the minimal model of a different theory. In fact, this seems to be realised by AdS/CFT, at least in the following simple example:  taking scalar field correlators for simplicity, the dictionary of \cite{Witten:1998qj,Gubser:1998bc} states that CFT correlators (given by a generating functional $Z_\text{CFT}[\phi_0]$) at a large $N$ limit are well-approximated by the (renormalised) on-shell bulk Euclidean AdS action $I_\text{AdS}[\phi]$ where $\phi_0$ is the boundary value of the on-shell bulk scalar field $\phi$:
\be
Z_\text{CFT}[\phi_0]=\exp(-I_\text{AdS}[\phi])\,,\qquad \phi \text{ solves }\frac{\delta I_\text{AdS}[\phi]}{\delta \phi}=0\,,\quad \phi|_{\partial\text{AdS}}=\phi_0\,.
\ee
The analogy to the Jevicki-Lee S-matrix prescription is obvious, modulo the need to discard the 2-point contribution on either side.

\bigskip

\noindent{\bf Acknowledgements:} It is a pleasure to acknowledge discussions and helpful feedback from Tim Adamo, Alec Barns--Graham, Panos Betzios, Jos\'e Eliel Camargo--Molina, Joe Keir,  Victor Lekeu, Iva Lovrekovic, Olga Papadoulaki, Matt Roberts, David Tennyson, and Dan Thomson. I also appreciate the clarifications of Tommaso Macrelli, Christian S\"amann, and Martin Wolf regarding their work \cite{Macrelli:2019afx}. Special thanks goes to Jim Stasheff for encouragement and for pointing out certain references I missed.

I am supported by the EPSRC programme grant ``New Geometric Structures from
String Theory'' (EP/K034456/1).

\appendix
\section{Cyclic Hodge-Kodaira decompositions}
We prove \eqref{hodgekodaira} and the claims around it, for finite $\dim \mathcal V$. Identity \eqref{cyclicity} for a cyclic inner product $\kappa_{ab}$ implies in particular for the unary bracket $K(T_a)=C^b_a T_b$
\be
\kappa_{ac}C^c_b=(-1)^{ab}\kappa_{bc} C^c_a\implies \kappa(v_1, Kv_2)=(-1)^{(\deg v_1)(\deg v_2)}\kappa(v_2,Kv_1)
\ee
whence
\be
(\im K)^\perp=\ker K\implies (\ker K)^\perp=\im K
\ee
(we need finite dimensionality for the last implication). Therefore $\kappa$ descends to a non-degenerate bilinear form on the the cohomology $\ker K/\im K$ of $K$.

Now select a subspace $P$ of cohomology representatives in $\ker K$ so that
\be
\ker K=P\oplus \im K\,.
\ee
We see that $\kappa|_{P\times P}$ is non-degenerate, implying
\be
\mathcal V=P\oplus P^\perp
\ee
where we stress $P\cap P^\perp=\{0\}$ and $\kappa$ is non-degenerate on both $P$ \emph{and} $P^\perp$.

Since $\im K \subset P^\perp$ and $\im K \subset (\im K)^\perp$ it follows that $\kappa$ restricts to zero on $\im K$ in $P^\perp$. For any complementary subspace $L$ so $P^\perp=\im K\oplus L$ is a direct sum, $K$ can be restricted to a map $K|_L:L\to\im K$ where it is invertible, and $L\cong \im K$ as vector spaces. Call the inverse $G:\im K\to L$ (which is necessarily a degree $-1$ linear map given $\deg K=+1$) and extend it to a map $\mathcal V\to L$ by setting $G(v)=0\quad\forall v\in P\oplus L\subset \mathcal V$. This in particular implies
\be
G^2=0
\ee
and we get the identities
\be
(KG)^2=KG\,,\quad (GK)^2=GK\implies GKG=G\,,\quad KGK=K\,.
\ee
Since $L$ and $\im K$ are both $\kappa$-perpendicular to $P$, if $P$ denotes the projector onto $P$ we also get $KP=PK=KG=GK=0$.

$KG$ and $GK$ are projectors onto $\im K$ and $L$ respectively, but \emph{not} $\kappa$-orthogonal ones since $\im K$ and $L$ can never be chosen to be orthogonal. However, we can select $L$ such that $\kappa|_{L\times L}=0$ i.e.~$\im(GK)$ is $\kappa$-null (easy to construct in finite dimensions). When this is the case, we obtain the identity
\be
\label{appendixGantisymmetry}
\kappa_{a c}G^c_b=-(-1)^{ab}\kappa_{bc}G^c_a\,.
\ee

To summarise: a ``propagator'' $G:\mathcal V\to\mathcal V$ of degree $-1$ compatible with $\kappa$ \eqref{appendixGantisymmetry} exists and depends on a choice of cohomology representatives $P$ and a choice of ``co-exact'' elements $L\subset P^\perp$. This is true in finite dimensions. For a Hilbert space if we assume that $(\im K)$ and $(\ker K)$ are both closed the argument goes through \emph{except} for the point that we can always find an $L$ such that $\kappa|_{L\times L}=0$.

\section{Path integral derivation of the Zinn--Justin equation from the quantum master equation}
Using the definitions in the main text, since we only ever use left derivatives,
\be
\big(\Gamma,\Gamma\big)=2\int dx\;(-1)^\Phi \frac{\delta \Gamma}{\delta \Phi(x)}\frac{\delta \Gamma}{\delta\starred \Phi(x)}=2\int dx\; J(x)\frac{\delta \Gamma}{\delta\starred \Phi(x)}\,.
\ee
We need to calculate $\delta\Gamma/\delta\starred\Phi(x)$ in terms of $Z[J,\starred\Phi]$. This is in fact proportional to $-i\hbar\delta \log Z[J,\starred\Phi]/\delta \starred\Phi(x)|_{J=J[\Phi,\starred\Phi]}$ by the following short calculation: from the Legendre transform \eqref{legendre},
\begin{align}
\frac{\delta \Gamma}{\delta\starred \Phi(x)}&=-i\hbar\left(\frac{\delta}{\delta\starred\Phi(x)}\log Z\big[J[\Phi,\starred\Phi],\starred\Phi\big]-\int dy\; \frac{\delta J(y)}{\delta\starred\Phi(x)} \Phi(y)\right)\\
&=-i\hbar\left(\frac{\delta\log Z}{\delta\starred\Phi(x)}\Big|_{J=J[\Phi,\starred\Phi]} +\int dy\; \frac{\delta J(y)}{\delta \starred\Phi(x)}\frac{\delta \log Z}{\delta J(y)}  -\frac{\delta J(y)}{\delta\starred\Phi(x)} \Phi(y)             \right)
\end{align}
so the last two terms cancel by \eqref{Phidefinitionlegendre}. We used the chain rule for left derivatives $df=dw^a (\partial z^b/\partial w^a) (\partial f/\partial z^b)$ (the order is important when fermion variables are involved).

Therefore
\begin{align}
\big(\Gamma,\Gamma\big)=2 \int dx\;\frac{1}{Z[J,\starred\Phi]} \int \mathcal D\phi \;J(x)e^{\int J\phi} \frac{\delta}{\delta\starred \Phi(x)}e^{iS/\hbar} 
\end{align}
where $J=J[\Phi,\starred\Phi]$ and $S=S[\phi,\starred\Phi+\delta\Psi/\delta\phi]$ as in section 3. Using the trick $J(x)\exp(\int J \phi)=(-1)^\phi \delta \exp(\int J \phi)/\delta\phi(x)$, we can integrate by parts inside the path integral to move $\delta/\delta\phi(x)$ to $\exp(iS/\hbar)$ (at the expense of a total derivative term, which we discard). The end result is
 \begin{align}
\big(\Gamma,\Gamma\big)=-2 \int dx\;\frac{1}{Z[J,\starred\Phi]} \int \mathcal D\phi \;e^{\int J\phi} \Delta e^{iS/\hbar} 
\end{align}
(where the $x$ integral smears $\Delta=(-1)^\phi \delta^2/\delta\phi(x)\delta\starred\phi(x)$ in $x$) which leads to the result in the main text.

\bibliography{../NewBib}
\end{document}